\documentclass[prd,onecolumn,12pt,tightenlines,superscriptaddress,nofootinbib,showpacs,shownokeys]{revtex4} 

\usepackage{graphicx}
\usepackage{psfrag}

\begin{document}

\title{The uniformly most powerful test of statistical significance for 
counting-type experiments with background.}

\author{L.~Fleysher}
\email{lazar.fleysher@physics.nyu.edu}
\author{R.~Fleysher}
\email{roman.fleysher@physics.nyu.edu}
\affiliation{Department of Physics, New York University, New York,
                New York 10003}

\author{T.~J.~Haines}
\email{haines@lanl.gov}
\affiliation{Los Alamos National Laboratory, Los Alamos,
             New Mexico 87545}
\author{A.~I.~Mincer}
\email{allen.mincer@nyu.edu}
\author{P.~Nemethy}
\email{peter.nemethy@nyu.edu}
\affiliation{Department of Physics, New York University, New York,
                New York 10003}

\date{June 13, 2003}

\begin{abstract}
In this paper, after a discussion of general properties of statistical
tests, we present the construction of the most powerful hypothesis test
for determining the existence of a new phenomenon in counting-type
experiments where the observed Poisson process is subject to a Poisson
distributed background with unknown mean.
\end{abstract}


\maketitle

\section{Introduction}

Typical experiments which search for new phenomena such as
rare decays (see, for
example~\cite{Bdecay}), new particles (see, for
example~\cite{amanwimps}) and astronomical gamma-ray and X-ray sources
(see, for example~\cite{rosat,milagrito_grb}) are counting-type
experiments. In such experiments the number of observed events is
distributed according to a Poisson distribution with some average rate.
Unfortunately, often such experiments are subject to unwanted
background, i.e., even if the new phenomenon is not present, the
experiment will register some number of counts with average background
rate. Only in rare cases is the expected background rate known.
Therefore, to overcome this difficulty, these experiments typically
utilize one of several available techniques. One of the possibilities is
to perform two observations --- one for which some of the observed
counts are believed to originate from the new phenomenon and the other
for which all observed counts are known to originate due to background
only; all other conditions of the observations are kept intact. Thus,
the two observations will yield two observed counts $n_{1}$ and $n_{2}$
made during observation times $t_{1}$ and $t_{2}$ respectively. The
number of events $n_{1,2}$ in each observation is drawn from the
corresponding parent Poisson distribution. If the new phenomenon
exists, the observations will come from the Poisson distributions with
different average event rates. If, on contrary, the new phenomenon does
not exist, the observations will come from the Poisson distributions
with identical average event rates. When it is not possible to obtain
data due to background only or to otherwise determine the average
expected background rate another approach is often used: the first
observation is made as before, but the second one is made with the help
of computer simulations of exactly the same experiment with the new
phenomenon ``turned off''. In other words, the observation $n_{1}$
during time $t_{1}$ is obtained as in the previous case. The second
observation $n_{2}$ during time $t_{2}$ is obtained by simulating the
experiment using only the established laws of physics. Since it is
believed that the computer simulation correctly describes the data
collection procedure, $n_{2}$ can be regarded as drawn from the Poisson
distribution with average background rate. In either case, a decision as
to the plausibility of the existence of the phenomenon is made based on
the outcomes of the two observations. Because the outcomes of the
observations represent random numbers drawn from their respective parent
distributions, the question of existence of the new phenomenon is here
addressed by a hypothesis test.

Various statistical tests have been developed which address the question
of testing the hypothesis that two independent observations $n_{1}$ and
$n_{2}$ made during times $t_{1}$ and $t_{2}$ are due to common
background only. The methods, an overview of which can be found
in~\cite{zhang_ramsden}, mostly use Gaussian-type approximations to the
Poisson distribution and are not reliable for small numbers of observed
events. In this paper we present the construction of the most powerful
hypothesis test for this situation. That is, we calculate the critical
region to be used which, for any given probability of claiming
consistency with background fluctuation (typically a number such as
$10^{-3}$ or less), maximizes the probability of detecting a signal.

\section{Testing a statistical hypothesis}

\begin{figure}
\centering
\psfrag{yy}{$y$}
\psfrag{yc}{$y_{c}$}
\psfrag{pp}{$p(y)$}
\psfrag{p0}{$p_{0}(y)$}
\psfrag{p1}{$p_{1}(y)$}

\includegraphics[width=0.9\columnwidth]{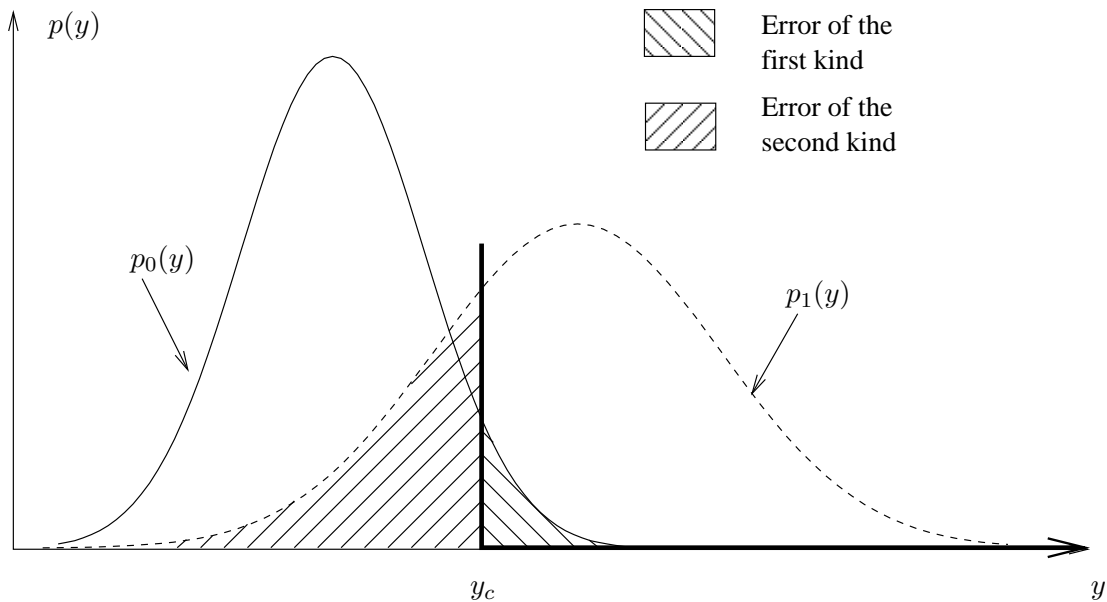}
\caption{\label{fig:errors} Illustration of a critical region with
         corresponding errors of the first and second kind.}
\end{figure}

We begin by reviewing the general procedure of hypotheses tests outlined
in~\cite{neyman_pearson}. A statistical hypothesis is a statement
concerning the distribution of a random variable $Y$.\footnote{The
variable $Y$ may be a multidimensional vector. Throughout, we will use
capital letters (such as $Y$) to denote random variables and small
letters (such as $y$) to denote their particular realizations.} A
hypothesis test is a rule for accepting or rejecting the hypothesis
based on the outcome of an experiment. The hypothesis being tested,
called the \textit{null hypothesis} $H_{0}$, is formulated in such a way
that all prior knowledge strongly supports it.  The hypothesis is
rejected if the observed value $y$ of the random variable $Y$ lies
within a certain critical region $w$ of the space $W$ of all possible
outcomes of $Y$ and accepted or doubted otherwise. It follows, then,
that if there are two tests for the same hypothesis $H_{0}$, the
difference between them consists in the difference in critical regions.
It also follows that $H_{0}$ can be rejected when, in fact, it is true
(\textit{error of the first kind}); or it can be accepted when some
other \textit{alternative hypothesis} $H_{1}$ is true (\textit{error of
the second kind}). The existence of an alternative hypothesis is clear,
otherwise the null hypothesis would not be questioned. The probabilities
of errors of these two kinds depend on the choice of the critical
region. The definitions are illustrated in figure~\ref{fig:errors}. Each
probability $p(y)$ of occurrence of every event $y$ is given a subscript
corresponding to its progenitor hypothesis, such as $p_{0}(y)$ for
$H_{0}$. The region to the right of $y_{c}$ is selected to be the
critical one and the two types of errors of the test are marked with
different hatching.

A critical region is said to be \textit{the best critical region} for
testing hypothesis $H_{0}$ with regard to $H_{1}$ if it is the one which
minimizes the probability of the error of the second kind (to accept
$H_{0}$ when $H_{1}$ is true) among all regions which give the same
fixed value of the probability of the error of the first kind (to reject
$H_{0}$ when it is true). The construction of the best critical region,
resulting in the most powerful test of $H_{0}$ with regard to $H_{1}$
was considered in \cite{neyman_pearson} where the problem is solved for
the general case of simple hypotheses. A hypothesis is said to be
\textit{simple} if it completely specifies the probability of the
outcome of the experiment; it is \textit{composite} if the probability
is given only up to some unspecified parameters. In general, if at least
one of the hypotheses is composite, the best critical region may not
exist \cite{neyman_pearson}.

Critical regions $w(\alpha)$ corresponding to different probabilities
$\alpha$ of errors of the first kind are engineered before the test is
performed. When experimental data $y$ is obtained, the smallest $\alpha$
is found such that $y \in w(\alpha)$. It is then said, that the observed
experimental data can be characterized by the $p$-value equal to
$\alpha$. 

The maximum $p$-value at which the null hypothesis is rejected is called
\textit{significance} of the test and will be denoted as $\alpha_{c}$. 
The corresponding critical region will be denoted as
$w_{c}=w(\alpha_{c})$.  The significance $\alpha_{c}$ is set in advance,
before the test is performed and its choice is based on the penalty for
making the error of the first kind. (False scientific discoveries should
not happen very often, and thus the significance is often selected as
$\alpha_{c}=10^{-3}$.) One minus the probability of the error of the
second kind is called the \textit{power of the test} which we denote as
$(1-\beta)$. 

If as the result of the experiment the observed data lies inside of the
critical region $w_{c}$, it is concluded that the null hypothesis is
rejected in favor of the alternative one with significance $\alpha_{c}$
and power $(1-\beta)$. If, however, the observed data lies outside of 
the critical region $w_{c}$, it is concluded that the null hypothesis is 
not rejected in favor of the alternative one with significance 
$\alpha_{c}$ and power $(1-\beta)$.

Special consideration must be given to the case of a composite null
hypothesis $H_{0}$ of the form $p_{0}(y; \{\lambda\})$ with unknown
parameters $\{\lambda\}$. Indeed, suppose a critical region $w$ is
specified.  Then, it is possible to perform the test: if the obtained
value of the observable $Y$ is inside the critical region $w$, the null
hypothesis is rejected, if $y \not\in w$, it is accepted. However, the
probability of the error of the first kind $\alpha(\{\lambda\}) =
\int_{w} p_{0}(y; \{\lambda\}) dy$ in general depends on unknown values
of parameters $\{\lambda\}$ of the null hypothesis. Thus, the $p$-value
$\alpha$ can not be assigned and the conclusion of the test can not be
stated. It is therefore desired to construct such critical regions $w$,
that the probability of the error of the first kind does not depend on
the values of unknown parameters. Such regions are called
\textit{similar to $W$ with regard to parameters $\{\lambda\}$}. A
method for construction of such regions was found in
\cite{neyman_pearson} under limited conditions which in the case of one
parameter $\lambda$ are:

\begin{itemize}
\item the probability distribution $p_{0}(y;\lambda)$ is infinitely
      differentiable with respect to $\lambda$ and
\item the probability distribution $p_{0}(y;\lambda)$ is such that if 
      $\Phi(y) = \frac{d \log p_{0}}{d\lambda}$, then

      \begin{equation}
      \frac{d \Phi}{d\lambda} = A + B \cdot \Phi
      \label{equation:neyman_pearson}
      \end{equation}

      where the coefficients $A$
      and $B$ are functions of $\lambda$, but are independent of
      observations $y$.
\end{itemize}

If the above conditions are satisfied, critical regions $w$ similar to 
$W$ with regard to $\lambda$ are built up of pieces of the hypersurfaces 
$\Phi = const$ defined by the likelihood ratio $p_{1}/p_{0}>q$.

\section{Null hypothesis being tested}

As was pointed out above, in a typical counting-type experiment two
independent observations yielding two counts $n_{1}$ and $n_{2}$ are
made during time periods $t_{1}$ and $t_{2}$ respectively with all other
conditions being equal. Because it is assumed that each event carries no
information about another, each of the observed counts can be regarded
as being drawn from a Poisson distribution with some value of the
parameter. In as much as an attempt is being made to establish the
existence of a new phenomenon, the null hypothesis $H_{0}$ is formulated
as: $n_{1}$ and $n_{2}$ constitute an independent sample of size 2 from
a single Poisson distribution (adjusted for the duration of observation)
which is due to a common background with some unknown count rate
$\lambda$, or

\[
 H_{0}: \;\;\; 
    p_{0}(n_{1}, n_{2}) = \frac{(\lambda t_{1})^{n_{1}}}{n_{1}!}
                          \frac{(\lambda t_{2})^{n_{2}}}{n_{2}!}
                           e^{- \lambda (t_{1} + t_{2})}
\]

The alternative hypothesis is that the two observations are due to
Poisson distributions with different unknown count rates $\lambda_{1}$
and $\lambda_{2}$ respectively ($\lambda_{1} \ne \lambda_{2}$):

\[
 H_{1}: \;\;\;
    p_{1}(n_{1}, n_{2}) = \frac{(\lambda_{1} t_{1})^{n_{1}}}{n_{1}!}
                           e^{- \lambda_{1} t_{1}}
                          \frac{(\lambda_{2} t_{2})^{n_{2}}}{n_{2}!}
                           e^{- \lambda_{2} t_{2}}
\]

The usual physical situation is that one of the count rates is
considered to have some amount due to the new process and the remainder
due to the background process [e.g. $\lambda_{1} = (\lambda +
\lambda_{signal})$ and $\lambda_{2}=\lambda$], thus the formulated
hypothesis test matches the physical problem given in the introduction.

It is seen that for the case of interest both hypotheses are of
composite type ($\lambda$'s are unspecified) which complicates the
construction of the test.

\section{The most powerful test}

The formulated probability distribution $p_{0}(n_{1}, n_{2})$ satisfies
the conditions of a special case considered in \cite{neyman_pearson},
which facilitates the search for the best critical region in $(n_{1},
n_{2})$ space. 

The probability distribution $p_{0}(n_{1}, n_{2})$ satisfies the following
conditions:

\begin{itemize}
\item $p_{0}(n_{1}, n_{2})$ is infinitely differentiable with respect to
      $\lambda$, 
\item the function $\Phi(n_{1}, n_{2})$ defined as 
      $\Phi(n_{1}, n_{2}) = \frac{d \log p_{0}}{d\lambda}$ satisfies
      equation (\ref{equation:neyman_pearson}) with
      $\Phi(n_{1},n_{2}) = \frac{n_{1}+n_{2}}{\lambda}-(t_{1}+t_{2})$, 
      $A=-\frac{t_{1}+t_{2}}{\lambda}$ and
      $B=-\frac{1}{\lambda}$.
\end{itemize}

Therefore, the best critical region corresponding to the error of the
first kind $\alpha$, determined from $\Phi =const$, is built up of
pieces of the lines $n_{t}=n_{1}+n_{2}=const$. The segments of each line
are those for which the ratio of likelihoods $p_{1}/p_{0}$ is greater
than some constant $q_{\alpha}$. This translates to:

\[
  \left(\frac{\lambda_{1}}{\lambda}\right)^{n_{1}} 
  \left(\frac{\lambda_{2}}{\lambda}\right)^{n_{t}-n_{1}}
                    e^{-(\lambda_{1}-\lambda)t_{1}}
                    e^{-(\lambda_{2}-\lambda)t_{2}} \ge q_{\alpha}
          \ \ \ \Rightarrow \ \ \ 
       n_{1}\ln\frac{\lambda_{1}}{\lambda_{2}}\ge q^{\prime}_{\alpha} \]
which can be written as:

\begin{equation}
  n_{1} \ge n_{\alpha},    \;\;\; \lambda_{1} > \lambda_{2}
\label{equation:critical:best:1}
\end{equation}

\begin{equation}
  n_{1} \le n_{\alpha},    \;\;\; \lambda_{1} < \lambda_{2}
\label{equation:critical:best:2}
\end{equation}

where the critical value $n_{\alpha}$ is chosen to satisfy the desired
probability of the error of the first kind $\alpha$:

\[
  \alpha \sum_{k=0}^{n_{t}} p_{0}(k,n_{t}-k) = 
      \sum_{k=n_{\alpha}}^{n_{t}} p_{0}(k,n_{t}-k), 
                           \;\;\; \lambda_{1} > \lambda_{2}
\]
\[
  \alpha \sum_{k=0}^{n_{t}} p_{0}(k,n_{t}-k) = 
      \sum_{k=0}^{n_{\alpha}} p_{0}(k,n_{t}-k), 
                           \;\;\; \lambda_{1} < \lambda_{2}
\]

Substituting explicitly the expression for $p_{0}(n_{1},n_{2})$ in to
these equations, we obtain

\begin{equation}
  \alpha = (1+\gamma)^{-n_{t}}\sum_{k=n_{\alpha}}^{n_{t}} 
  C_{n_{t}}^{k} \; \gamma^{k}=
  I_{\frac{\gamma}{1+\gamma}}(n_{\alpha},n_{t}-n_{\alpha}+1),
  \;\;\;\;\; \lambda_{1} > \lambda_{2}
\label{equation:signif:best:1}
\end{equation}

\begin{equation}
  \alpha = (1+\gamma)^{-n_{t}} \sum_{k=0}^{n_{\alpha}} 
  C_{n_{t}}^{k} \; \gamma^{k}=
  I_{\frac{1}{1+\gamma}}(n_{t}-n_{\alpha},n_{\alpha}+1),
  \;\;\;\;\; \lambda_{1} < \lambda_{2}
\label{equation:signif:best:2}
\end{equation}

where $\gamma = t_{1}/t_{2} > 0$, $C_{n}^{m} = \frac{n!}{m!(n-m)!}$ are
binomial coefficients and $I_{x}(a,b)$ is the normalized incomplete beta
function. It must be emphasized that the critical value $n_{\alpha}$
depends on the parameters $\lambda_{1,2}$ of the alternative hypothesis
only via the relation $\lambda_{1} < \lambda_{2}$ or $\lambda_{1} >
\lambda_{2}$. The best critical region for testing the null hypothesis
against the alternative with $\lambda_{1} \ne \lambda_{2}$ does not
exist, but it does exist for testing against $\lambda_{1} > \lambda_{2}$
or $\lambda_{1} < \lambda_{2}$ separately, that is when the signal is a
source or a sink respectively. The equations
(\ref{equation:signif:best:1}) with (\ref{equation:critical:best:1}) or
(\ref{equation:signif:best:2}) with (\ref{equation:critical:best:2})
define the best critical region $w(\alpha)$ in the space $(n_{1},n_{2})$ 
for testing $H_{0}$ with regard to $H_{1}$ (defined for 
$\lambda_{1}<\lambda_{2}$ and $\lambda_{1}>\lambda_{2}$ separately)
corresponding to the probability of the error of the fist kind $\alpha$.
The boundary of this critical region is found by solving the equation
(\ref{equation:signif:best:1}) or (\ref{equation:signif:best:2}) with
respect to $n_{\alpha}$ for all possible values of $n_{t}$. Owing to the
discrete nature of the observed number of events these equations might
not have solutions for the specified level of significance $\alpha_{c}$.
Nevertheless, it is possible to construct a \textit{conservative}
critical region such that the probability to observe data within the
region does not exceed the preset level $\alpha_{c}$ if the null
hypothesis is true. This is done by requiring:

\[ \left\{\begin{array}{l}
  \alpha_{c} \geq 
        I_{\frac{\gamma}{1+\gamma}}(n_{\alpha},n_{t}-n_{\alpha}+1) \\
  \alpha_{c} <
        I_{\frac{\gamma}{1+\gamma}}(n_{\alpha}-1,n_{t}-n_{\alpha}+2) \\
  \end{array}\right.
  \;\;\;\;\; \lambda_{1} > \lambda_{2} \]

\[ \left\{\begin{array}{l}
  \alpha_{c} \geq
         I_{\frac{1}{1+\gamma}}(n_{t}-n_{\alpha},n_{\alpha}+1) \\
  \alpha_{c} <                                         
         I_{\frac{1}{1+\gamma}}(n_{t}-n_{\alpha}-1,n_{\alpha}+2)
  \end{array}\right.
  \;\;\;\;\; \lambda_{1} < \lambda_{2} \]

The power $(1-\beta)$ of the test will, of course, depend on the values
of the parameters of the alternative hypothesis:

\[ (1-\beta)=
    \sum_{n_{t}=0}^{\infty}\sum_{k=n_{\alpha}}^{n_{t}} p_{1}(k,n_{t}-k),
                                \ \ \  \lambda_{1} > \lambda_{2} \]
\[ (1-\beta)=
    \sum_{n_{t}=0}^{\infty}\sum_{k=0}^{n_{\alpha}} p_{1}(k,n_{t}-k),
                                \ \ \ \lambda_{1} < \lambda_{2}  \]
After explicit substitution of $p_{1}(n_{1},n_{2})$ into these 
equations, we obtain:


\begin{equation}
  (1-\beta)=\sum_{n_{t}=0}^{\infty}
          \frac{(\lambda_{1}t_{1}+\lambda_{2}t_{2})^{n_{t}}}{n_{t}!}
                             e^{-(\lambda_{1}t_{1}+\lambda_{2}t_{2})}\,
   I_{\frac{\lambda_{1}t_{1}}{\lambda_{1}t_{1}+\lambda_{2}t_{2}}}(n_{\alpha},n_{t}-n_{\alpha}+1),
                                      \ \ \ \lambda_{1} > \lambda_{2}
\label{equation:power:best:1}
\end{equation}


\begin{equation}
   (1-\beta)=\sum_{n_{t}=0}^{\infty}
          \frac{(\lambda_{1}t_{1}+\lambda_{2}t_{2})^{n_{t}}}{n_{t}!}
                             e^{-(\lambda_{1}t_{1}+\lambda_{2}t_{2})}\,
   I_{\frac{\lambda_{2}t_{2}}{\lambda_{1}t_{1}+\lambda_{2}t_{2}}}(n_{t}-n_{\alpha},n_{\alpha}+1),
                                      \ \ \ \lambda_{1} < \lambda_{2}
\label{equation:power:best:2}
\end{equation}

For the purposes of the hypothesis test itself, equations
(\ref{equation:signif:best:1},\ref{equation:signif:best:2}) provide the
method for the $p$-value calculation without the need for solving them.
To do this, $n_{t}$ must be set to $(n_{1}+n_{2})$ and $n_{\alpha}$ to
$n_{1}$ then $\alpha$ is computed from the equation
(\ref{equation:signif:best:1}) or (\ref{equation:signif:best:2}). If the
obtained $p$-value $\alpha$ is not greater than $\alpha_{c}$ the null
hypothesis is rejected. This is the uniformly most powerful test of
$H_{0}$ with regard to $H_{1}$.

It can be seen that the application of the method of best critical
region construction~\cite{neyman_pearson} to the problem of testing
whether two observations came from the Poisson distributions with the
same parameter or not have led us to the criterion suggested on
intuitive grounds in~\cite{przybor}. The presented discussion, however,
shows that it is not possible to construct a better test for the
hypotheses under consideration. The practical use of equations
(\ref{equation:signif:best:1},\ref{equation:signif:best:2},%
\ref{equation:power:best:1},\ref{equation:power:best:2}) should not
present any difficulty using modern computers~\cite{recipes}.

\section{Compounding Results of Independent Tests}

It is often the case that the complete data set consists of several runs
of the experiment, each of which belongs to the counting type subjected
to Poisson distributed background with unknown means. The data set is
then a set of pairs $(n_{1,r},n_{2,r})$ with corresponding durations of
observations $(t_{1,r},t_{2,r})$, where subscript $r$ enumerates all the
runs of the experiment. Here we distinguish two cases: first where the
parameters of both hypotheses do not depend on the run number $r$ and
second where such independence can not be asserted because of
modifications made to the apparatus between the runs. In the first case
it can be seen that the critical region must be constructed out of
pieces of surfaces $n_{t} = \sum_{r} (n_{1,r} + n_{2,r}) = const$, on
which $\sum_{r} n_{1,r} \ge n_{\alpha}$ for the case of $\lambda_{1} >
\lambda_{2}$ or $\sum_{r} n_{1,r} \le n_{\alpha}$ for the case of
$\lambda_{1} < \lambda_{2}$. It is thus seen that equations
(\ref{equation:signif:best:1},\ref{equation:signif:best:2}) provide a
method for the $p$-value calculation: $n_{t} = \sum_{r} (n_{1,r} +
n_{2,r})$, $n_{\alpha} = \sum_{r} n_{1,r}$ and $\gamma = \frac{\sum_{r}
t_{1,r}}{\sum_{r} t_{2,r}}$. In other words, if the parameters of both
hypotheses do not depend on the run number $r$, the corresponding
observations can simply be added. 

In the second case, the derivation proceeds in the fashion similar to
the presented derivation of equations
(\ref{equation:signif:best:1},\ref{equation:signif:best:2}), the
critical region is built up of surfaces $n_{t,r} = n_{1,r} + n_{2,r} =
const$ such that

\[
  \sum_{r} n_{1,r} \log (\lambda_{1,r} / \lambda_{2,r}) \ge 
                                                    q^{\prime}_{\alpha}
\]

The critical value $q^{\prime}_{\alpha}$ is chosen to satisfy the
desired probability of the error of the first kind $\alpha$:

\[
 \alpha =
        \prod_{r} (1+ \gamma_{r})^{-n_{t,r}}
        \sum_{ \left\{ \begin{array}{l}
                       k_{r} \in [0, n_{t,r}] \\
                       \sum_{r} k_{r} \log (\lambda_{1,r} / \lambda_{2,r})
                                 \ge q^{\prime}_{\alpha}
                       \end{array}
               \right.
   } C_{n_{t,r}}^{k_{r}} \gamma_{r}^{k_{r}}
\]

The $p$-value is obtained by setting $q^{\prime}_{\alpha}$ equal to
$\sum_{r} n_{1,r} \log (\lambda_{1,r} / \lambda_{2,r})$. In this case,
the critical region depends on the parameters of the alternative
hypothesis, but the test becomes uniformly most powerful if it is known
that the ratios $\lambda_{1,r} / \lambda_{2,r}$ do not depend on the run
number $r$. The latter situation is common in practice because it
reflects change of the level of pre-scaling of events in the data
acquisition system or degradation of efficiency of sensors occurred
between the experimental runs.

\section{Conclusion}

In this paper we have reviewed the basic concepts of statistical
hypothesis tests and underlined the relevant aspects often employed. The
difficulty arises because frequently both the null and the alternative
hypotheses are of composite type.

We have considered typical counting experiments and constructed the most
powerful statistical test. In doing so, we have insisted on the ability
to quantify the error of the first kind although the parameter of the
composite null hypothesis is unknown. The test also happens to be the
uniformly most powerful with regard to the composite alternative
hypothesis with $\lambda_{1}<\lambda_{2}$ or $\lambda_{1}>\lambda_{2}$
separately. The constructed test is especially important for the case of
small number of events where previously used methods are inadequate
because the usual Gaussian-type approximations break down. Fortuitously,
this is the case for which the proposed test can most easily be
performed. The existence of the most powerful statistical test allows
comparisons with other computationally less demanding methods to be made
which may be important for some applications.

\begin{acknowledgments}

We would like to thank Prof. James Linnemann for helpful discussions and
for making us aware of some relevant references.

This work is supported by the National Science Foundation (Grant Numbers
PHY-9901496 and PHY-0206656), the U. S. Department of Energy Office of
High Energy Physics, the Los Alamos National Laboratory LDRD program and
the Los Alamos National Laboratory Institute of Nuclear and Particle
Astrophysics and Cosmology of the University of California.

\end{acknowledgments}

\bibliography{poisson_paper}
\bibliographystyle{plain}








\end{document}